# Time resolved measurements on low-density single quantum dots at 1300nm

C. Zinoni[*,1], B. Alloing[1], C. Monat[1], L. H. Li[1], L. Lunghi[2], A. Gerardino[2] and A. Fiore[1,2]

[1] Ecole Polytechnique Fédérale de Lausanne (EPFL), CH-1015 Lausanne, Switzerland
[2] Institute of Photonics and Nanotechnology, CNR, Rome, Italy.

We present time integrated and time resolved measurements on single In/As quantum dots (QD) emitting at 1300nm, at 10K, embedded in a planar microcavity. We clearly identify a standard spectroscopic signature from single QDs and compare the exciton line width and biexciton (BX) binding energy for several QDs. We present this data for QDs spatially selected by etched mesas and metallic apertures. Time resolved photoluminescence (PL) from single electron-hole recombination in the ground state of the QD is investigated as a function of excitation power and temperature. The measurements reveal the presence of a background emission that delays the PL of excitons in the ground state.

**1 Introduction** Efficient generation of single photons on demand at telecom wavelength (1310nm and 1550nm) is crucial for quantum key distribution (QKD). The optical properties of single quantum dots (QDs) have the potential for satisfying the requirements of a convenient single photon source: QDs can be grown in conventional semiconductor epitaxial systems and the nature of the 3D confinement of the wavefunction generates atomic-like spectral features. A single QD can be populated by several electron-hole pairs (excitons), each recombining to emit a photon. Due to Coulomb interactions originating from the strong charge confinement, the energy levels are shifted and each transition can be spectrally isolated to produce a single photon source. QDs that emit in the spectral range where silicon technology is used have been extensively studied [1-3]. However, the attenuation in optical fibers below 1270nm restricts the potential use of these QDs in QKD applications. Working with QDs emitting in the telecom window presents several challenges: first the QD emission has to be red shifted while maintaining a low spatial density—a difficult combination of requirements for conventional epitaxial growth methods. Second, the single photon detection technology for the near infrared is still in its infancy: noise levels, quantum efficiency, and temporal response are considerably poorer when compared to the single photon detection modules operating below 1000nm. Recently we have demonstrated [4] a technique for achieving ultra low areal densities (1–2 dots/µm2) and large dot sizes for emission in the 1300nm band. In combination with ultra small light-emitting diode structures [5-6], they could lead to electrically pumped single photon sources. These QDs present several distinct features, as compared to widely studied short-wavelength QDs, such as larger confinement energy, higher strain fields, and consequently different electron-hole wave functions. Time-resolved studies on single exciton transitions provide important information on the dynamics of the electron-hole relaxation into these new QDs and may change the way in which QD based single photon devices are operated. In this work we compare the photoluminescence from the ground state of the QDs spatially isolated by two different methods: etched mesas and metallic apertures. We also investigate the time evolution of the BX emission as a function of excitation power and temperature. We reveal the presence of a signal that delays the standard BX emission.

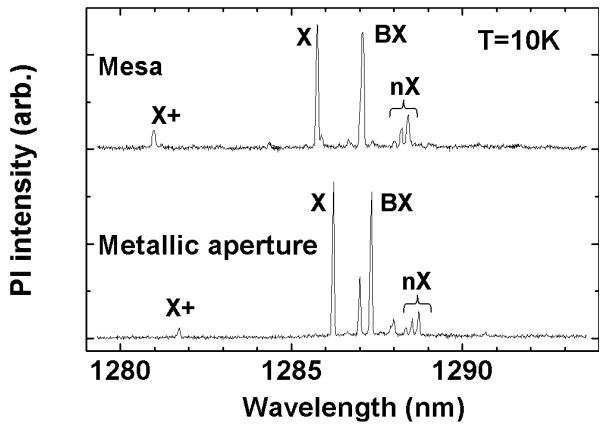

**Fig. 1** Photoluminescence excitation of two single quantum dots spatially isolated by a mesa and by a metallic aperture.

**2 Experimental** The low density InAs/GaAs QDs (2 dots/ $\mu m^2$) were grown by molecular beam epitaxy using an ultra-low growth rate technique of which a detailed description can be found in Ref. 4. Two techniques were adopted to spatially separate single QDs: apertures of 1 $\mu m^2$ in gold masks obtained by lift-off using a chemically amplified resist (UVIII) patterned by 100kV e-beam lithography, and 1 $\mu m^2$ etched mesas realized by reactive-ion-etching using a $SiO_2$ mask patterned by e-beam lithography. A pulsed diode laser emitting at 750 nm, with a maximum repetition rate of 80 MHz, was focused down to a 4µm spot on the sample with a microscope objective (numerical aperture 0.5). The photoluminescence was collected with the same objective and dispersed into a 1 m focal length monochromator equipped with a cooled InGaAs photodiode array detector; the spectral resolution of the setup is better than 30µeV (0.04nm). The photoluminescence collected by the objective can also be coupled into a single mode fiber and used for time-resolved PL measurements. A fiber coupled bandpass filter [tunable between 1270 and 1310nm with a full width at half maximum (FWHM) of 0.8nm] is used for selecting a single excitonic emission, while an InGaAs APD (idQuantique) detector operated in Geiger mode is used for detection.

**3 Results and Discussion** In Fig.1 we compare the spectra of the ground state of two single QDs emitting at the same energy. In both devices (metallic apertures and mesas) we observe the same spectral signature. First, two emission lines dominate the spectrum: the exciton (X) and biexciton (BX). As previously demonstrated [4-9], the identification follows from the power dependence of the integrated PL intensity. On average the exciton intensity is proportional to $P^{0.7\pm0.1}$ and the BX to $P^{1.4\pm0.1}$, where P is the pump power and the ratio of 2 between the coefficients is characteristic for X-BX dynamics. Secondly, the positive trion is always located at approximately 5nm from the exciton on the high energy side. Finally, the group of three lines marked nX (which are attributed to the recombination of single excitons on the ground state in the presence of charges in the excited state). This group of lines is always red-shifted with respect to the X and BX transitions by an amount comparable to the BX binding energy. The spectral line located between the X and BX of the QD isolated in a metallic aperture (Fig.1) is not always present in the spectra and it is tentatively attributed to the negative trion. No correlation was found between the presence of the negative exciton and the method used to spatially isolate the QD. In Fig.2a we plot the distribution of the line width of the exciton as a function of recombination energy for QDs isolated in mesas and metallic apertures. The emission range spreads over 30meV and does not influence the line width of the exciton: this is an indication of the deep confinement in these quantum dots, as compared to smaller QDs emitting below 1000nm, where the WL is a few meV from the ground state in the QDs. In Fig. 2b we show that the BX binding energy is not correlated to the X emission energy, which is usually the case for QDs emitting in the visible.

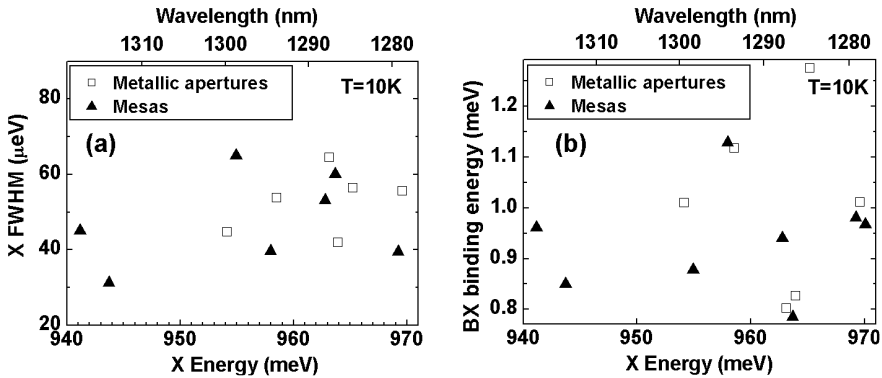

**Fig. 2** Statistics on single quantum dots spectral features. (a) FWHM of exciton vs. recombination energy. (b) BX binding energy vs. exciton recombination energy.

By coupling the QD photoluminescence into a single mode fiber and tuning the bandpass filter, we measured the lifetime of the biexciton for different pump powers and temperatures. For these measurements the InGaAs APDs were biased above avalanche threshold for 100 ns at a repetition rate of 1 MHz with a dead time of 10 µs. Due to the mismatch between the APDs spectral response and laser wavelength, the setup resolution was measured with a sample of GaNInAs quantum wells, emitting at 1300 nm at room temperature, with a lifetime previously measured to be 50 ps [10], which is well below the temporal resolution of the detector (600 ps). Fig.3 shows the time resolved PL of the BX emission at different pump powers, where the PL intensities are normalized to the integration time. At low excitation intensity the time evolution of the PL shows a fast rise limited by the resolution of the setup. A slight delay is observed before the decay of the signal from the BX which is usually attributed to population of the excited states. In a previous study [9] the BX decay was fitted by convoluting the setup response function with a one time constant exponential decay function to give a lifetime of 1.0±0.1ns. As the pump power in increased to 750pW to saturate the BX emission, the delay in the decay of the PL increases to 1.5ns. An additional increase in the excitation intensity further shifts the decay of the BX to 4ns, exposing another decay developing on faster time scales (limited by the resolution of the setup) at the same recombination energy of the BX. This new decay curve will be referred to as the *fast emission*. The same behavior is observed on the X emission, the positive trion and the ensemble of dots. The tail in the PL, at long delay times for higher power arises from the repopulation of the BX state

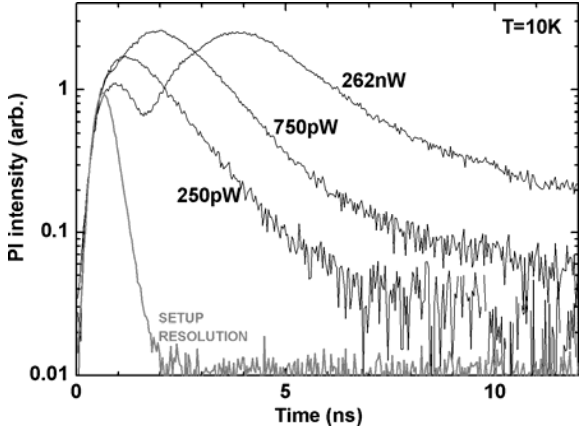

**Fig. 3** Time-resolved PL on BX emission from a single QD for increasing excitation intensity.

during the same excitation cycle from the higher lying states. To isolate and identify the fast emission, we increased the temperature of the sample to 55K and compared the photoluminescence decay at several positions in the broadened spectrum of the QD (Fig.4). The first measurement was taken at 1288 nm, corresponding to the edge of the QD ground state emission, the second at 1295 nm to include a single exciton transition, and finally at 1305 nm to measure the decay of the broad background typical of QDs at these temperatures. The measurements were made at high excitation intensities (230nW). The time resolved PL is shown in Fig. 4b, where we clearly isolate the fast emission (at 1288 nm) and measure a deconvoluted lifetime of 0.7 ns. The measurements at 1295 nm and 1305 nm present similar characteristics, they are both delayed by the fast emission and have similar lifetimes. The exciton emission at 1295 nm appears to be delayed by the broad background. From the evidence collected it would appear that the emission from the ground state (GS) is characterized by distinct phases. In an initial phase, after the pump pulse, a large density of carriers surrounding the QD is strongly interacting with excitons in the QD, so that the Coulomb interactions between carriers in the ground state are not anymore the dominant terms in determining the excitonic configuration. As a consequence, we obtain a broad band emission with a rise time limited by the resolution of the APD. As the number of carriers in the wetting layer is reduced by radiative, non-radiative recombination and repopulation of the dot,

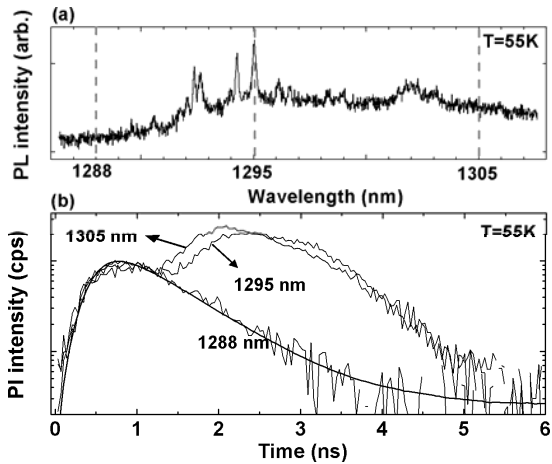

**Fig. 4** (a) PL signal from the same QD as in Fig. 3 at 55K. Dashed lines represent the setting of the band-pass filter for the time-resolved measurements shown in (b). The lifetime measurement at 1288nm is shown with the least squares fit.

the delicate balance between the Coulomb terms is restored and we observe the expected decay from the individual exciton transitions in the QD. The effect of external carriers has been previously observed in the PL properties, e.g. in the jitter of the spectral lines of single excitons [9]. At low excitation intensity the amount of charges in the wetting layer is smaller and the fast emission is less significant, as observed in our measurements. The low QD density (1dot/$\mu m^2$) could be responsible for amplifying such a phenomenon: a large amount of QDs per unit surface are capable of depleting the carriers in the wetting layer on a faster time scale as compared to low density dots, thus limiting the fast emission to shorter time scales and weaker intensities. It is interesting to note that the fast emission is a source of uncorrelated photons which will limit the ability of QDs to operate as sources of triggered single photons: pumping resonantly in the excited state of deeply confined

QDs eliminates the presence of carriers in the WL, thus reducing the interaction of excitons in the QD with external carriers. It has been show that resonant excitation improves the $g^{(2)}(0)$ by an order of magnitude [10].

**4 Conclusion** We have shown that our QDs have very repeatable spectral features. Neither the line width of the exciton transition nor the BX binding energy show a clean correlation with the emission energy, which is probably due to the combination of composition and size variations. Time resolved measurements on single quantum dots reveal the presence of a fast decay component that could be characteristic of low density QD samples. Resonant excitation could be a solution to avoid non correlated photon emission during antibunching experiments.

**References**

[1] M. D. H. J. Kimble and L. Mandel, Phys. Rev. Lett. **39**, 691 (1977).
[2] T. Basché, W. E. Moerner, M. Orrit, and H. Talon, Phys. Rev. Lett. **69**,1516 (1992).
[3] C. Kurtsiefer, S. Mayer, P. Zarda, and H. Weinfurter, Phys. Rev. Lett. **85**, 290 (2000).
[4] B. Alloing, C. Zinoni, V. Zwiller, L. H. Li, C. Monat, M. Gobet, G. Buchs, A. Fiore, E. Pelucchi, and E. Kapon, Appl. Phys. Lett. **86**, 101908 (2005).
[5] A. Fiore J. X. Chen and M. Ilegems , Appl. Phys. Lett. **81**, 1756 (2002).
[6] C. Zinoni, B. Alloing, C. Paranthoën, and A. Fiore , Appl. Phys. Lett. **85**, 2178 (2002).
[7] C. Zinoni, B. Alloing, C. Monat, V. Zwiller, L. H. Li, A. Fiore, L. Lunghi, A. Gerardino, H. de Riedmatten, H. Zbinden, and N. Gisin, Appl. Phys. Lett. **88**, 131102 (2006).
[8] A. Markus, A. Fiore, J. D. Ganière, U. Oesterle, J. X. Chen, B. Deveaud, M. Ilegems, and H. Riechert, Appl. Phys. Lett. **80**, 911 (2002).
[9] B. Patton, W. Langbein, and U. Woggon, Phys. Rev. B **68**, 125316 (2003).
[10] A. Bennett, D. Unitt, P. Atkinson, D. Ritchie, and A. Shields, Opt. Express **13**, 50 (2005).